\documentclass[10 pt,conference, letterpaper]{IEEEtran}

\usepackage{fancyhdr, epsfig, epsf, amsthm, amsmath, amssymb, amsfonts, subfigure, color}
\usepackage{threeparttable,dsfont,enumerate,comment,multirow,bigstrut}
\usepackage[noadjust]{cite}
\newtheorem{theorem}{Theorem}

\newtheorem{lemma}{Lemma}
\newtheorem{corollary}{Corollary}

\newtheorem{proposition}{Proposition}

\def\qed{$\blacksquare$}

\def\endproof{\hfill \qed}

\def\E{\mathsf{E}}

\def\SIR{\mathsf{SIR}}

\def\l{\left}
\def\r{\right}
\def\({\left(}
\def\){\right)}
\def\lb{\left\{}

\def\[{\left[}
\def\]{\right]}

\def\lb{\textsf{LB}}
\def\ub{\textsf{UB}}
\def\figwidth{8.8}

\def\papertitle{Coverage and Rate of Downlink Sequence Transmissions with Reliability Guarantees}
\IEEEoverridecommandlockouts

\begin{document}
\title{ \fontsize{20}{22}\selectfont  \papertitle}

\author{Jihong~Park and Petar~Popovski 
\thanks{J.~Park and P.~Popovski are with Department of Electronic Systems, Aalborg University, Denmark (email: \{jihong, petarp\}@es.aau.dk).}
\thanks{This work has been in part supported by the European Research Council (ERC Consolidator Grant Nr. 648382 WILLOW) within the Horizon 2020 Program, and partly by Innovation Fund Denmark via the Virtuoso project.}
}
\maketitle 

\begin{abstract} Real-time distributed control is a promising application of 5G in which communication links should satisfy certain reliability guarantees. In this letter, we derive closed-form maximum average rate when a device (e.g. industrial machine) downloads a sequence of $n$ operational commands through cellular connection, while guaranteeing a certain signal-to-interference ratio ($\SIR$) coverage for all $n$ messages. The result is based on novel closed-form $n$-successive $\SIR$ coverage bounds. The proposed bounds provide simple  approximations that are increasingly accurate in the high reliability region.
\end{abstract}

\section{Motivation and Contribution}

One of the focus in the emerging 5G wireless systems will be communication with reliability guarantees. The most challenging variant is ultra-reliable low-latency communication (URLLC), with extreme reliability guarantees.
A representative use case for URLLC is industrial Internet of Things (IoT) \cite{EricssonIIoT:15}, in which a certain user equipment (UE), such as an actuator, is controlled in real time by downloading operational commands for high-precision cloud manufacturing \cite{5GPPPfactory:15}. A loss of a command sequence may cause critical malfunction or shutdown of the process. Therefore, the network should guarantee wireless coverage with very high reliability, e.g. $\geq 99\% $ in order to decode the $n$ successive commands. 

In this letter we rely on stochastic geometry to derive $n$-successive $\SIR$ (Signal-to-Interference Ratio) coverage $p_n(t)$ for a target $\SIR$ threshold $t$ and $n\geq 1$, defined as $p_n(t):=\Pr(\SIR_1\geq t, \SIR_2\geq t, \cdots, \SIR_n\geq t)$. We thereby derive the closed-form maximum average rate $\mathcal{R}_n(\eta)$ that ensures $\eta$ reliability for $n$-successive message receptions, i.e. $p_n(t)\geq \eta$. Regardless of whether the reliability is extreme as in URLLC or more relaxed, the introduction of a reliability guarantee poses a challenging problem. This precludes us from directly exploiting previously known closed-form $\SIR$ coverage and rate expressions \cite{Andrews:2011bg,Haenggi:ISIT14,JHParkTWC:15}. The reason comes from the following two issues: \emph{spatial correlation} during $n$-successive receptions and the \emph{trade-off between coverage and rate}.

\begin{figure}\label{Fig:Network}
	\centering
	\includegraphics[width=\figwidth cm]{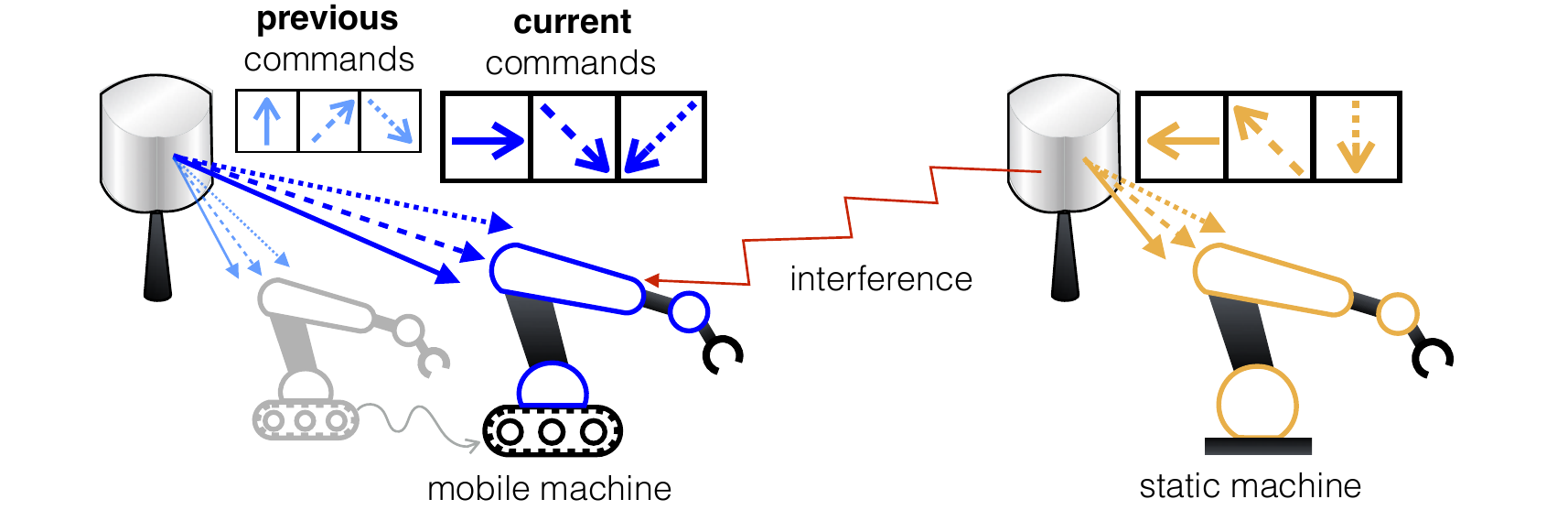}
	\caption{A sequence of $3$ command receptions for mobile and static UEs.}
\end{figure}

Regarding spatial correlation, if UE locations are independent and identically distributed (i.i.d.) during $n$ message receptions, then $p_n(t) = p_1(t)^n$ straightforwardly.\footnotemark\; However, a URLLC UE should receive $n$ messages within a short time span, during which all UEs and the interferers can be considered static, see Fig. 1. For a cellular downlink with fixed UE locations, the $n$-successive $\SIR$ coverage has been known as (Theorem 2, \cite{Haenggi:ISIT14}):
\begin{align} \label{Eq:Cov_exact}
p_n(t)=1/{}_2F_1\(n, -2/\alpha; 1-2/\alpha; -t\)
\end{align}
at a randomly picked (i.e. a typical) UE.\footnotemark\; The denominator ${}_2F_1(a, b;c;z)$ is a Gauss hypergeometric function that can only be numerically computed, defined as ${}_2F_1(a, b;c;z):=\sum_{n=0}^\infty \frac{\Gamma(a + n)\Gamma(b+n)\Gamma(c)}{\Gamma(a)\Gamma(b)\Gamma(c+n)} \frac{z^{n}}{n!}$ where $\Gamma(x)$ denotes the gamma function. It is thus difficult to establish a direct relation between $n$ and $p_n(t)$ due to the analytical intractability of the hypergeometric function. 

\addtocounter{footnote}{-1}\footnotetext{For a typical UE's $\SIR$, i.i.d. mobility is almost surely identical to any Markovian mobility such as a random walk and L\'evy flight thanks to displacement theorem \cite{HaenggiSG} as utilized in \cite{MobilMFGSG:GC16} (see \cite{Dhillon:16} for general mobility).}
\stepcounter{footnote}\footnotetext{A special case is $p_1(t) = 1/\;_2F_1\(1, -\frac{2}{\alpha}; 1-\frac{2}{\alpha}; -t \)$ in \cite{Haenggi:ISIT14}, which is identical to $1/(1 + t^{2/\alpha} \int_{u>  t^{-2/\alpha}} du/[1+u^{\alpha/2}] )$ presented in \cite{Andrews:2011bg}.}

\begin{figure*}
\centering
 	\subfigure[for the range within $\[0,1\]$.]{\includegraphics[width=9 cm]{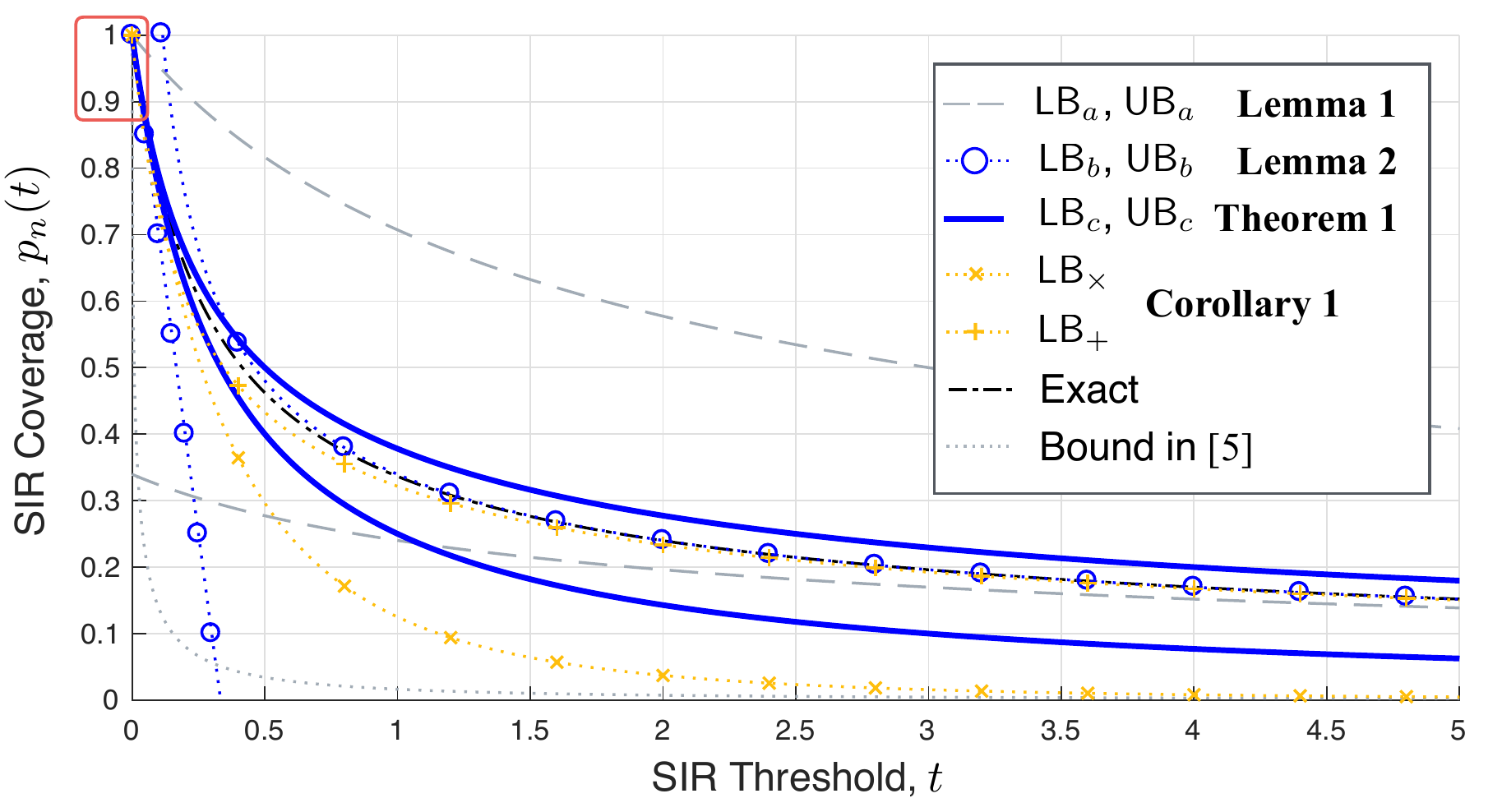}} 
 	\subfigure[for the range within $\[0.9,1\]$.]{\includegraphics[width=9 cm]{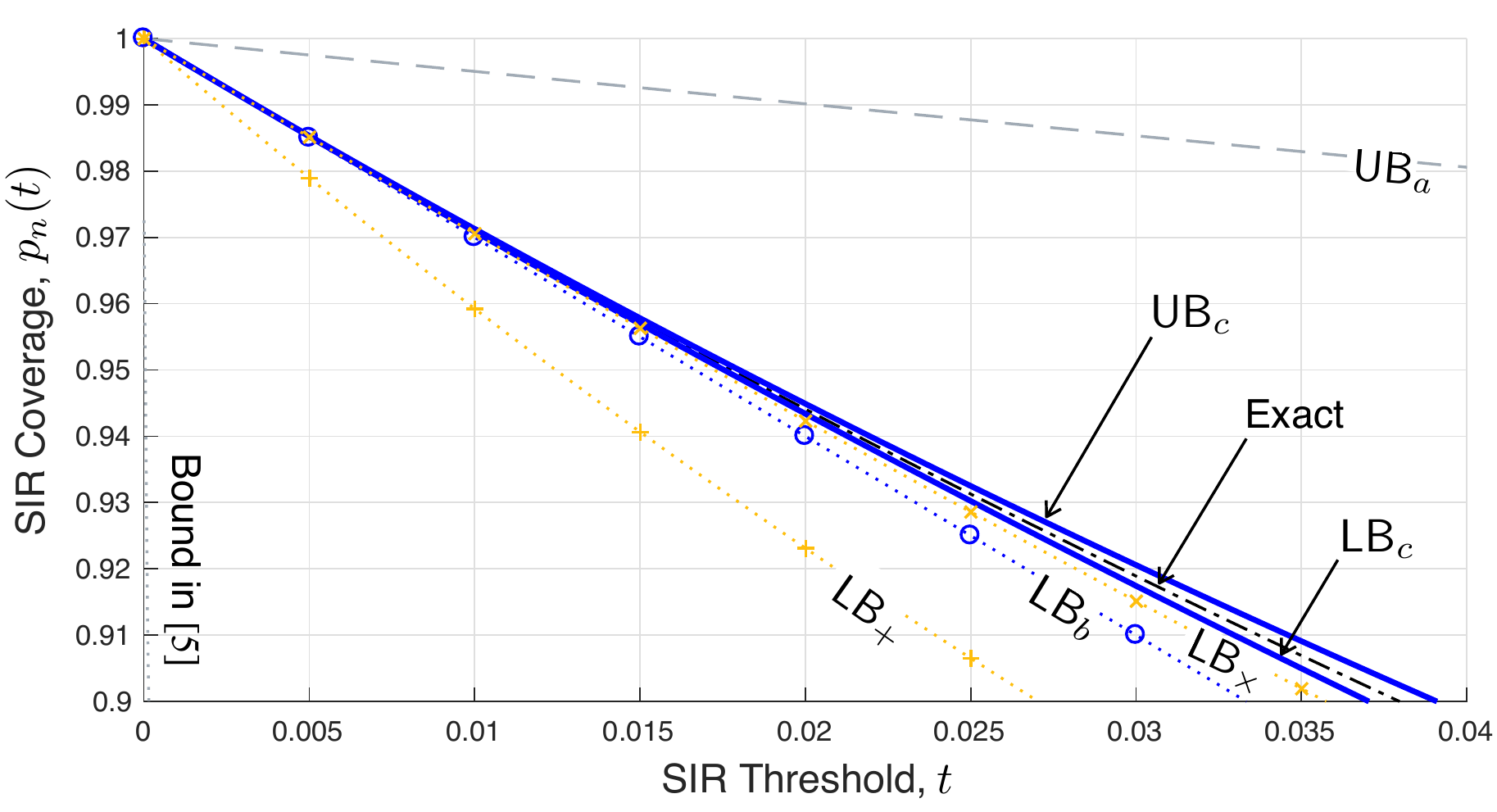}}
	\caption{$3$-successive $\SIR$ coverage bounds (\textbf{Theorem~1}, $\alpha=4$), derived via the bounds after Pfaff transformation (\textbf{Lemma 1}) and the  asymptotic bounds (\textbf{Lemma 2}). They are compared to the existing bounds $1/(1 + {c_n} t^{2/\alpha})$ in \cite{JHParkTWC:15} and ${c_n}^{-1}t^{-2/\alpha}$ ($=\textsf{UB}_a$) in \cite{PoissLattice:13,Dhillon:12}. The exact curve follows from \eqref{Eq:Cov_exact}.}
\end{figure*}

Next, the average rate expression $\E[\log(1 + \SIR)]$ in previous works \cite{Andrews:2011bg,Haenggi:ISIT14,JHParkTWC:15} bore no relationship to the reliability constraint $p_n(t)\geq \eta$. The reason is because in those works it was implicitly assumed that each BS knows the full channel state information (CSI) of all the desired and interfering links. This enables the BS to adaptively adjust its modulation and coding scheme (MCS) such that $t=\SIR$ resulting in $p_n(\SIR)=1$. The full CSI unfortunately cannot be exchanged within a low target latency, and is therefore not affordable for URLLC. 

Without CSI, the $\SIR$ threshold is kept constant during $n$ transmissions, while satisfying $p_n(t)\geq \eta$. This induces the trade-off between a typical UE's coverage $p_n(t)$ and its quasi-concave average rate $p_n(t)\log(1 + t)$ with respect to $t$. We focus on the maximum average rate $\mathcal{R}_n(\eta)$ given as
\begin{align}
\mathcal{R}_n(\eta):= \underset{{t\geq 0}}{\arg\max}\; p_n(t)\log(1 + t)\quad \text{s.t.}\; p_n(t)\geq \eta.
\end{align}
Although (2) is a simple convex optimizaition with an inequality condition, it cannot be analytically solved due to the intractable hypergeometric function of $p_n(t)$ in (1).


A suitable way to avoid the difficulties brought by the hypergeometric function is to utilize more tractable $\SIR$ coverage bounds, as done in our recent work \cite{UR2Cspaswin:17}. Nonetheless, the bounds in~\cite{UR2Cspaswin:17} are limited only to the case $n=1$. Other existing $\SIR$ coverage bounds \cite{PoissLattice:13,Dhillon:12, JHParkTWC:15} are too loose for ultra-reliable scenarios, within the range $[0.9, 1]$ for low $\SIR$ threshold $t$, as Fig. 2 shows.


In this letter we propose novel closed-form bounds of $n$-successive $\SIR$ coverage $p_n(t)$ (see $\textsf{LB}_c$ and $\textsf{UB}_c$ in \textbf{Theorem~1}), and thereby derive the closed-form bounds of maximum rate $\mathcal{R}_n(\eta)$ guaranteeing $p_n(t)\geq \eta$ (\textbf{Proposition 1}). For high reliability $\eta\geq 0.9$, the closed-form results show that the impact of spatial UE correlation during $n$ receptions is negligibly small. In this case the proposed lower bound under i.i.d. UE locations tightly approximates the exact rate ($\textsf{LB}_\times$ in \textbf{Corollaries 1} and~\textbf{2}). For moderate reliability $0.3<\eta<0.9$, a slightly modified lower bound becomes accurate as an approximation ($\textsf{LB}_+$ in \textbf{Corollaries 1} and \textbf{2}).

\section{System Model}
Consider a downlink cellular network in a two-dimensional plane. The network comprises uniformly located: BSs with density $\lambda_b$ and UEs with density $\lambda_u$. This leads to two independent homogeneous Poisson point processes $\Phi(\lambda_b)$ and $\Phi(\lambda_u)$, representing BS and UE locations. Each UE associates with its nearest BS, and requests $n$ messages. All messages are of unit size.
UE locations are fixed during the $n$-successive message receptions, and randomly and independently changed for the next $n$-successive message receptions. 

Each BS serves a single UE with unit transmission power and frequency bandwidth per time slot. We assume that $\lambda_u \gg \lambda_b$ such that any BS has at least a single associated UE to be served. Transmitted signals experience path loss attenuation with the exponent $\alpha>2$ and Rayleigh fading. A typical UE is located at the origin, capturing any locations without $\SIR$ changes according to Slyvnyak's theorem \cite{HaenggiSG}. The $i$-th message, $1 \leq i \leq n$, at the typical UE is received with 
$\SIR_i:=g^i r^{-\alpha}/\sum_{r_j>r} g_j^i {r_j}^{-\alpha}$, where $r_j:=|x_j|$ denotes the $j$-th nearest interfering BS distance for its location $x_j \in \Phi(\lambda_b)$. Fading power $g_j^i$ at the $i$-th reception is exponentially distributed with unit mean. We drop the subscript $j$ for the desired signal link. 
When $i$ changes from $1$ to $n$, the fading powers $g^i$ and $g_j^i$ vary with each $i$, while the desired BS and interferer distances $r$ and $r_j$ are fixed. In this case, $p_n(t)$ has been known as (1) according to Theorem 2 in \cite{Haenggi:ISIT14}.

\section{$n$-Successive $\SIR$ Coverage Bounds }
In this section we propose closed-form bounds of $n$-successive $\SIR$ coverage $p_n(t)$. The bounds are not only much tighter for URLLC but also as simple as the existing bounds. The desired bounds are derived by utilizing a transformed function of $p_n(t)$ and $p_n(t)$'s asypmtotic curves as follows.

We define the lower and the upped bounds of $p_n(t)$ through $p_n^{\lb_i}(t) \leq p_n(t) \leq p_n^{\ub_i}(t)$ for $i\in\{a, b, c, \times, +\}$. We then apply a hypergeometric function transform \cite{GEAndrews:Book} to (1), yielding the following lemma.
\begin{lemma} (Pfaff Transformed Bounds) At a static typical UE, $p_n(t)$ has the following lower and upper bounds.
\begin{align}
p_n^{\emph{\lb}_a}(t) &= {c_n}^{-1}(1 + t)^{-\frac{2}{\alpha}}\\
p_n^{\emph{\ub}_a}(t) &= (1 + t)^{-\frac{2}{\alpha} }
\end{align}
where $c_n:= \Gamma\(1-\frac{2}{\alpha}\)\Gamma\(n + \frac{2}{\alpha}\)/\Gamma(n)$\\
\emph{\begin{proof} 
By using Pfaff's hypergeometric function transform \cite{GEAndrews:Book}, ${}_2F_1\(n,-\frac{2}{\alpha};1-\frac{2}{\alpha},-t\) = (1+t)^{\frac{2}{\alpha}} G(\alpha,n,t)$ where $G(\alpha,n,t):={}_2F_1\(1-\[\frac{2}{\alpha}+n\],-\frac{2}{\alpha};1-\frac{2}{\alpha};\frac{t}{1+t}\)$ that has the range $[1,c_n]$. Applying this to \eqref{Eq:Cov_exact} leads to the result.
\end{proof}}
\end{lemma}
The term $c_n\geq 1$ is a monotone increasing/decreasing function of $n$ and $\alpha$ respectively. Its minimum value is thus $c_1=\frac{2\pi}{\alpha} \csc\(\frac{2\pi}{\alpha}\)\overset{\alpha\rightarrow\infty}{=}1$.

It is remarkable that both bounds identically share the simple term $(1+t)^{-\frac{2}{\alpha}}$ only with the constant multiplication difference. Nevertheless, they have the following limitations. First, $p_n^{\lb_a}(t)$ has the limited range $[0, {c_n}^{-1}]$ of which the maximum is much lower than our target minimum reliability $0.9$ under $2<\alpha\leq 6$ for all $n\geq 1$. Second, $p_n^{\ub_a}(t)$ is independent of $n$, and thus becomes less accurate for larger~$n$. 

For these reasons, we aim at further improving the bounds in Lemma~1 such that the bounds behave similar to the original $p_n(t)$ with respect to $n$ and $t$ under the entire range $[0,1]$ while converging to the asymptotic curve of $p_n(t)$ for $t\rightarrow~0$. To this end, we first derive the asymptotic bounds as follows.
\begin{lemma} (Asymptotic Bounds) At a static typical UE, $n$-successive $\SIR$ coverage $p_n(t)$ follows the asymptotic curves:
\begin{align}
p_n^{\emph{\lb}_{b}}(t) & =1-\frac{2n }{\alpha-2}t\\
p_n^{\emph{\ub}_b}(t) &= {c_n}^{-1}t^{-\frac{2}{\alpha}}.
\end{align}
\noindent{Proof:}
\emph{Applying Taylor expansion to (1) with the hypergeometric function definition, its first derivative satisfies: $\lim_{t\rightarrow 0} p_n(t)'=-(2 n t)/(\alpha-~2)$ and $\lim_{t\rightarrow \infty} p_n(t)'=t^{-\frac{2}{\alpha}}/c_n$. By the hypergeometric function definition, $p_n(0)=1$ and $\lim_{t\rightarrow \infty}p_n(t)=0$. The asymptotic curves thus become $p_n^{\lb_b}(t)$ and $p_n^{\ub_b}$ respectively as $t\rightarrow~0$ and $\infty$.}

\emph{To prove the asymptotic curves become the bounds of $p_n(t)$, we further identify the behavior of $p_n(t)$ by directly considering its first derivative given as: $p_n(t)'=-2 p_n(t)\[1 - (1+t)^{-n} p_n(t)\]/(\alpha t)$. This shows $p_n(t)'~<~0$ for all $t<\infty$. Since $p_n(t)$ is continuous by the hypergeometric function definition, its functional and derivative values at the extreme $t$'s indicate $p_n(t)$ is a function that is (strictly) decresing as well as (strictly) convex for all $t$ (strict conditions hold for $t<\infty$). The asymptotic curve $p_n^{\lb_b}(t)$ shares $p_n^{\lb_b}(0)=p_n(0)=1$ while having limited domain, thus becoming the lower bound. The curve $p_n^{\ub_b}(t)$ satisfies $\lim_{t\rightarrow\infty} p_n^{\ub_b}(t)=\lim_{t\rightarrow\infty} p_n(t)=0$ while having infinite maximum range as $t\rightarrow 0$, thus becoming the upper bound.} 
\endproof
\end{lemma}

Note that $p_n^{\ub_b}(t)$ in Lemma 2 generalizes its special case $p_1^{\ub_b}(t)$ that has been derived in \cite{PoissLattice:13,Dhillon:12}. It is also noted that the previous Lemma 1's $p_n^{\lb_a}(t)$ converges to the derived asymptotic curve $p_n^{\ub_b}(t)$ as $t\rightarrow \infty$.


Next, we combine Lemma 1 and $p_n^{\lb_b}(t)$ in Lemma 2, resulting in the following Theorem.
\begin{theorem} (Coverage Bounds) At a static typical UE, the $n$-successive $\SIR$ coverage $p_n(t)$ is bounded by:
\begin{align}
p_n^{{\emph{\lb}_c}}(t) &= \(1 + n t \)^{-\frac{2 }{\alpha-2}} \\
p_n^{{\emph{\ub}_c}}(t) &= \(1 + \frac{ n \alpha }{\alpha-2} t\)^{-\frac{2}{\alpha}}.
\end{align}
\noindent{Proof:}
\emph{We choose the common term $(1+t)^{-\frac{2}{\alpha}}$ in Lemma~1 as the baseline, which is also a strictly decreasing and convex function for $t<\infty$. We modify this, and aim at its convergence to the asymptotic curve $1-2 n t/(\alpha-2)$ in Lemma~2 as $t\rightarrow 0$. For this purpose, we consider its generalized function $(1+~A t)^{-\frac{2 B}{\alpha}}$ for constants $A, B>0$, and consider its Taylor expanded form $1-2 A B t/\alpha$ for an infinitesimal $t$, which should satisfy $A B = n\alpha/\(\alpha-2\)$ for the desired convergence. We select $\l\{A,B\r\}=\l\{n,\alpha/\(\alpha-2\)\r\}$ and $\l\{n\alpha/\(\alpha-2\),1\r\}$ respectively yielding $p_n^{\lb_c}$ and $p_n^{\ub_c}$.}

\emph{To verify the lower bound, showing $p_n(t)\geq p_n^{\lb_c}(t)$ only at a single point $t$ suffices thanks to their strictly decreasing and convex-shaped behaviors (see the proof of Lemma~2). In this respect, consider a sufficiently large constant $M>0$ such that $p_n^{\ub_b}(M)$ in Lemma~2 approximates $p_n(M)$. In this case, $p_n(M) \geq p_n^{\lb_c}(M) $ becomes $M^{\frac{2}{\alpha}}\geq {c_n}^{\frac{\alpha}{2}-1}/n$, which can always be satisfied by choosing a sufficiently large $M$.}

\emph{
Similarly, we prove the upper bound by showing $p_n^{\ub_c}(M)\geq p_n(M)\approx p_n^{\ub_b}(M)$. This is identical to proving $D(n,\alpha):=c_n\(\frac{\alpha-2}{n\alpha}\)^{\frac{2}{\alpha}}\geq 1$, which is a stricly increasing function of $n\geq 1$ and a quasi-concave function of $\alpha>2$ by the definition of $c_n$. Therefore, it suffices to verify $D(1,\alpha)= \( 1-\frac{2}{\alpha}\)^{\frac{2}{\alpha}}c_1 \geq 1$ as $\alpha$ goes to either $\infty$ or $2$ as follows. Firstly, recall $c_1\overset{\alpha\rightarrow \infty}{=}1$ discussed after Lemma 1, thereby yielding $\lim_{\alpha\rightarrow \infty}D(1,\alpha)= 1$. Secondly, $\lim_{\alpha\rightarrow 2}D(1,\alpha) \overset{(a)}{=}\lim_{\alpha\rightarrow 2} \(1-\frac{2}{\alpha}\)\Gamma\(1-\frac{2}{\alpha}\)\Gamma\(1 + \frac{2}{\alpha}\)\overset{(b)}{=}\lim_{\alpha\rightarrow 2}\Gamma\(2-\frac{2}{\alpha}\)\Gamma\(1+\frac{2}{\alpha}\)=1$, where $(a)$ and $(b)$ follow from $\lim_{\alpha\rightarrow 2}\(1-\frac{2}{\alpha}\)^{\frac{2}{\alpha}}/\(1-\frac{2}{\alpha}\)=1$ and $\Gamma(x+1)=x\Gamma(x)$. The results indicate that the minimum of $D(n,\alpha)$ is unity, and consequently $D(n,\alpha)\geq 1$ for all $\alpha>2$ and $n\geq 1$.
}
\endproof
\end{theorem}

As Fig. 2 illustrates, $p_n^{\lb_c}$ and $p_n^{\ub_c}$ have the full range $[0, 1]$, and are even tighter than $p_n^{\lb_b}$ within the range $[0.9, 1]$.

Along the same lines, we additionally consider different choices of $\{A,B\}$ leading to the following lower bounds.
\begin{corollary} At a static typical UE, $p_n(t)$'s lower bounds are:
\begin{align}
p_n^{{\emph{\lb}_{\times}}}(t) &=  \(1 +  t \)^{-\frac{2 n}{\alpha-2}} \\
p_n^{{\emph{\lb}_{+}}}(t) &= \(1 + {c_n}^{\frac{\alpha}{2}} t \)^{-\frac{2}{\alpha}}.
\end{align}
\noindent{Proof:}\emph{ Applying $\{A,B\} = \{1,n\alpha/(\alpha-2)\}$ and $\{{c_n}^{\frac{\alpha}{2}},1\}$ to the proof of Theorem 1 results in $p_n^{{\lb_{\times}}}(t)$ and $p_n^{{\lb_{+}}}(t)$.}
\end{corollary}
The first bound $p_n^{\lb_\times}(t)$ is tight under high reliabilty, i.e. low $t$ as Fig.~2 shows. In addition, it satisfies $p_n^{\lb_\times}(t) = p_1^{\lb_\times}(t)^n$. This equality indicates that static (LHS) and i.i.d. (RHS) UE locations during $n$ receptions identically affect $\SIR$ almost surely. Knowing that $p_n^{\lb_\times}(t)~\leq~p_n(t)$, we conclude \emph{spatial UE correlation during $n$ receptions increases successive $\SIR$ coverage}, compared to the coverage under i.i.d. mobility. Such a \emph{spatial correlation gain diminishes under high reliability}.


The second bound $p_n^{\lb_+}(t)$ accuracy increases with $t$ as Fig.~2 shows, which may complement Theorem~1 for less reliable applications. This is achieved by utilizing the opposite asymptotic curve $p_n^{\ub_b}(t)$ in Lemma 2 for high $t$ accuracy, instead of $p_n^{\lb_b}(t)$ exploited in Theorem 1.



\begin{figure*}
\centering
 	\subfigure[for the reliability $\eta$ within $\[0.3,1\]$.]{\includegraphics[width=8.6 cm]{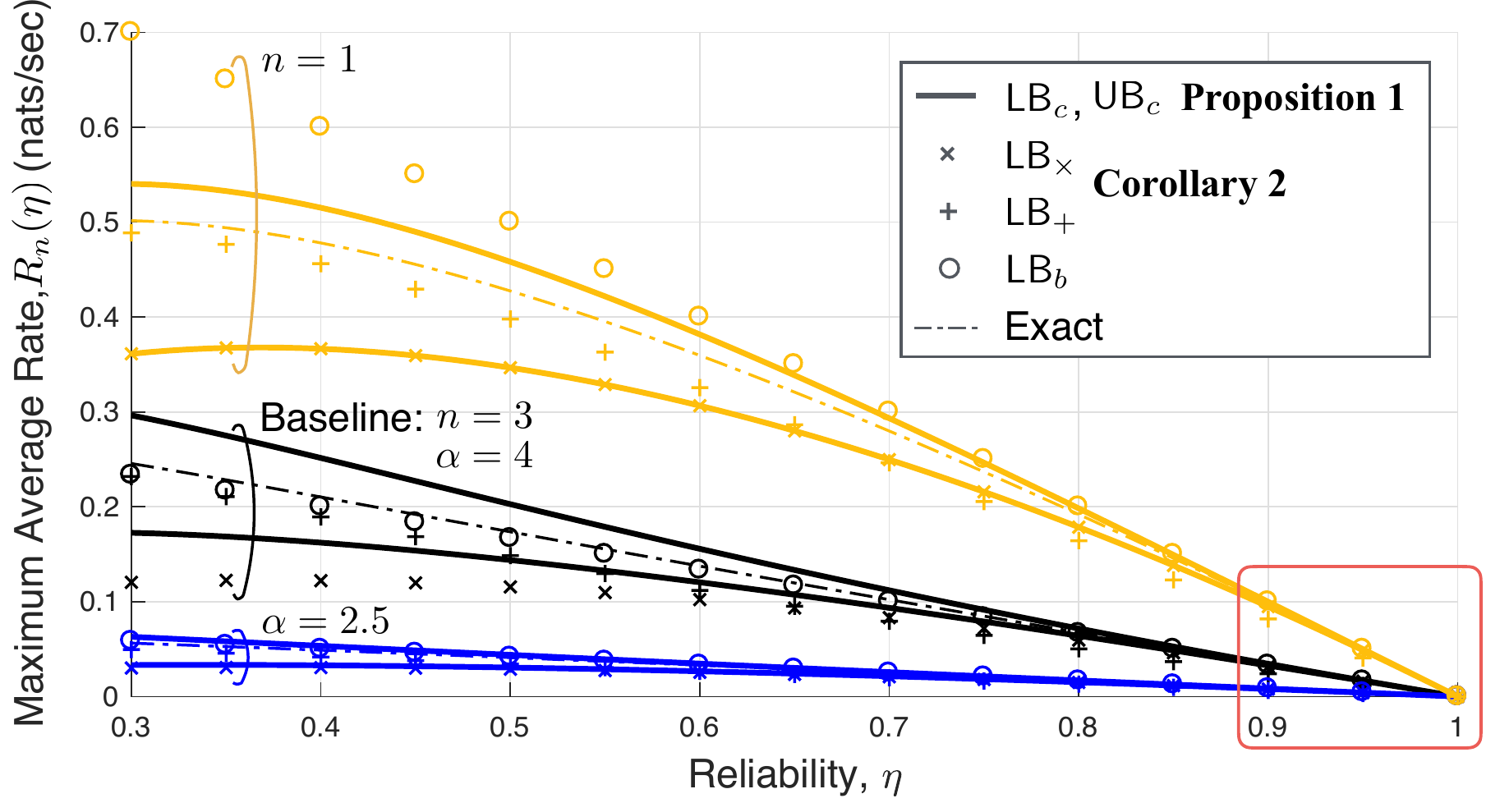}} 
 	\subfigure[for the reliability $\eta$ within $\[0.9,1\]$.]{\includegraphics[width=8.6 cm]{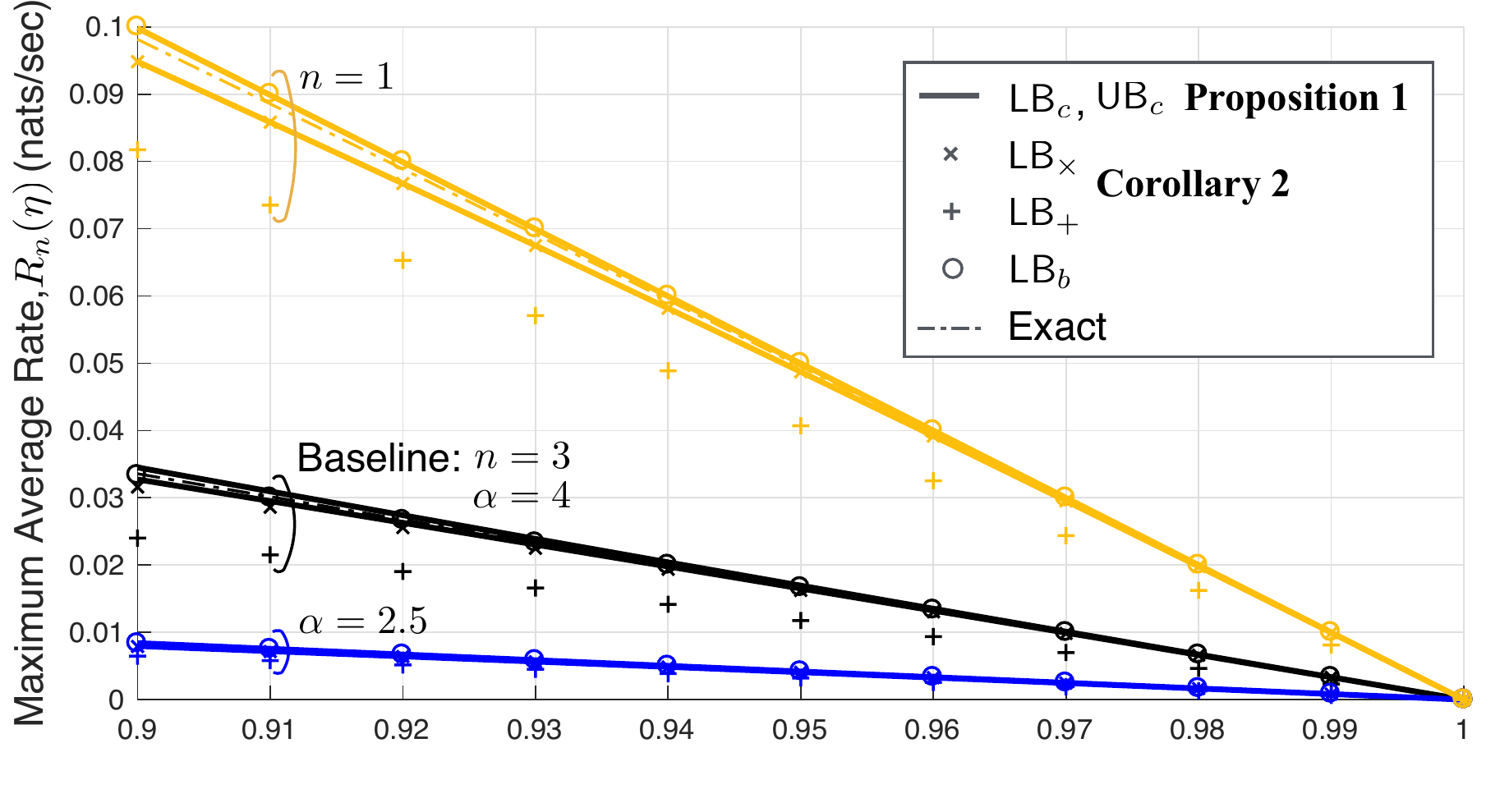}}
	\vspace{-5pt}\caption{The bounds of maximum average rate $\mathcal{R}_n(\eta)$ ensuring $n$-successive coveraege with probability $\eta$, under different $n$ and the path loss exponent $\alpha>2$.}
\end{figure*}

\section{Maximum Average Rate Bounds under $n$-Successive $\SIR$ Coverage Constraints}
In this section we derive the closed-form $\mathcal{R}_n(\eta)$, a typical UE's maximum average rate ensuring $p_n(t)\geq \eta$. For brevity, 
rate is measured in nats/sec ($1 \text{ nat} = 1/\log 2 \text{ bits}$).

Recall the optimization (2). We define $\hat{t}$ as the $t$ value when the reliability constraint's equality holds, i.e. $p_n(\hat{t})=\eta$. We consider $\hat{t}$ is the optimal $\SIR$ threshold maximizing average rate, i.e. $\mathcal{R}_n(\eta)=p_n\(\hat{t}\)\log(1 + \hat{t})$. To show this, we additionally define $\tilde{t}$ as the $t$ maximizing $\mathcal{R}_n(\eta)$ without the reliable constraint, i.e. $\tilde{t}:=\arg\max_{t\geq 0} p_n(t)\log(1+t)$. If $\hat{t}\leq \tilde{t}$, then the quasi-concave objective function strictly increases, having the maximum at $t=\hat{t}$. This is true for sufficiently high reliability, e.g. when $\eta\geq 0.29$ for all $n\geq 1$ and $2<\alpha\leq 4$, which is much less than our target $\eta\geq 0.9$.


For this reason, $\mathcal{R}_n(\eta)$ derivation is identical to deriving~$\hat{t}$. For URLLC, it can be analytically derived via Theorem 1. The original $\mathcal{R}_n(\eta)$ is then bounded by their corresponding results $\mathcal{R}_n^{\lb_i}(\eta) \leq \mathcal{R}_n(\eta) \leq \mathcal{R}_n^{\ub_i}(\eta)$ given as follows.
\begin{proposition} (Rate Bounds) At a static typical UE, maximum average rate $\mathcal{R}_n(\eta)$ satisfying $p_n(t)\geq \eta$ is bounded by:
\begin{align}
\mathcal{R}_n^{\emph{\lb}_c}(\eta) &=  \eta \log\(1 + \frac{\eta^{-\(\frac{\alpha}{2}-1\)}-1}{n} \)  \label{Eq:RateLB_woAMC}\\
\mathcal{R}_n^{\emph{\ub}_c}(\eta) &= \eta \log\( 1 + \[1-\frac{2}{\alpha}\]\frac{\eta^{-\frac{\alpha}{2}}-1}{n}\).
\end{align}
\emph{\begin{proof}
The optimal $\SIR$ threshold $\hat{t}$ is derived by taking the inverse $p_n(t)$ into $\eta$, i.e. $\hat{t} = p_n^{-1}(\eta)$. Applying $p_n^{\lb_c}$ and $p_n^{\ub_c}$ in Theorem 1 completes the proof.
\end{proof}}
\end{proposition}


Exploiting Corollary~1, we also derive the following bounds.
\begin{corollary} At a static typical UE, $\mathcal{R}_n(\eta)$'s lower bounds are:
\begin{align}
\mathcal{R}_n^{\emph{\lb}_\times}(\eta) &= \eta \log\( \eta^{-\frac{1}{n}\(\frac{\alpha}{2}-1\)} \) \\
\mathcal{R}_n^{\emph{\lb}_+}(\eta) &= \eta \log\( 1 + {c_n}^{-\frac{2}{\alpha}}\[\eta^{-\frac{\alpha}{2}}-1\]\).
\end{align}
\end{corollary}
As reliability increases, Fig.~3 shows that the exact $\mathcal{R}_n(\eta)$ curve from \eqref{Eq:Cov_exact} asymptotically converges to $R_n^{\lb_c}(\eta)$, $R_n^{\ub_c}(\eta)$, and $R_n^{\ub_\times}(\eta)$, as well as $R_n^{\lb_b}(\eta)=\frac{\alpha-2}{2n}(1-\eta)$ similarly derived by applying $p_n^{\lb_b}(t)$ in Lemma~2. Especially, the convergence to $R_n^{\ub_\times}(\eta)$ implies the spatial UE correlation impact becomes negligible in URLLC as stated after Corollary~1.

For moderate reliability $0.3<\eta<0.9$, $R_n^{\lb_+}(\eta)$ becomes accurate. Note that $\mathcal{R}_n^{\lb_b}(\eta)$ can also be an approximation if $n$ is large; otherwise, a huge gap exists as shown by the curve in Fig.~3 when $n=1$ (yellow circle). 

Additionally, it is noted that the proposed closed-form bounds are also applicable for conventional average rate calculation under the full CSI assumption as in \cite{Andrews:2011bg,Haenggi:ISIT14,JHParkTWC:15}. For example, when $n=1$, average rate is $\E[\log(1 + \SIR)] = \int_{0}^\infty \Pr\(\log(1 + \SIR)>x\) dx= \int_0^\infty  p_1\(e^{x}-1\) dx $. Applying the simplest bounds $p_1^{\lb_c}$ and $p_1^{\ub_a}$ to this yields the average rate bounds $(\alpha-2)/2$ and $\alpha/2$ respectively. Their linear combination with the equal weight factor leads to the rate approximation $(\alpha-1)/2$. For $\alpha=4$, this corresponds to $1.5$ nats/sec. The result is fit well with the exact calculation $1.49$ nats/sec while simplifying the known expression $\int_{x>0} dx/\( 1 + \sqrt{e^x-1}\[\pi/2 - \arctan(1/\sqrt{e^x-1}) \]\) $ in \cite{Andrews:2011bg}.

\section{Conclusion}
We proposed closed-form $n$-successive $\SIR$ coverage bounds, and thereby derived the closed-form maximum average rate guaranteeing such a successive coverage requirement. 


\bibliographystyle{ieeetr}  

\end{document}